\documentclass[]{article}
\usepackage{graphicx}
\usepackage{amsfonts,amsbsy,amscd,amsgen,dsfont,amsmath,amssymb,mathrsfs}
\usepackage{newtxtext}
\usepackage{newtxmath}
\usepackage{hyperref}
\hypersetup{
	colorlinks = true,
	urlcolor   = blue,
	citecolor  = black,
}
\usepackage{enumitem}
\usepackage{authblk}
\usepackage{lipsum}
\newcommand{\vc}{\mathbf}

\newcommand{\pard}[2]{\frac{\partial #1}{\partial #2}}

\newcommand{\gr}{\mbox{g}}

\begin{document}
	\title{Extreme firebrand transport by atmospheric waves in wildfires}
	\author[]{Mohammad Farazmand\thanks{Corresponding author's email address: farazmand@ncsu.edu}}
%	\email{farazmand@ncsu.edu}
	\affil{Department of Mathematics, North Carolina State University,
		2311 Stinson Drive, Raleigh, NC 27695-8205, USA}
	\date{}
	
	\maketitle
	
\begin{abstract}
In wildfires, burning pieces of ember---firebrands---are carried downstream by wind. At the time of landing, these firebrands can start secondary fires far away from the main burning unit. This phenomenon is called spotting and the secondary fires are referred to as spot fires. Here, we first present numerical evidence that atmospheric traveling waves can increase the spotting distance by at least an order of magnitude compared to unidirectional wind conditions. We then present theoretical results explaining this numerical observation. In particular, we show that the firebrand's motion can synchronize with the traveling wave, leading to a surf-like motion for some firebrand particles. This delays the firebrand's landing, making extreme spotting distances possible. This physical phenomena helps explain the discrepancy between previous theoretical estimates of maximum spotting distance and much larger spotting distances observed empirically. We derive new analytical expressions for the landing time and landing distance of the firebrands.
\end{abstract}
%\begin{keywords}
%	Wildfire; Transport; Traveling waves; Rare Events;
%\end{keywords}

%{\bf MSC Codes }  {\it(Optional)} Please enter your MSC Codes here

   \section{Introduction}
\begin{figure}
	\centering
	\includegraphics[width=0.9\textwidth]{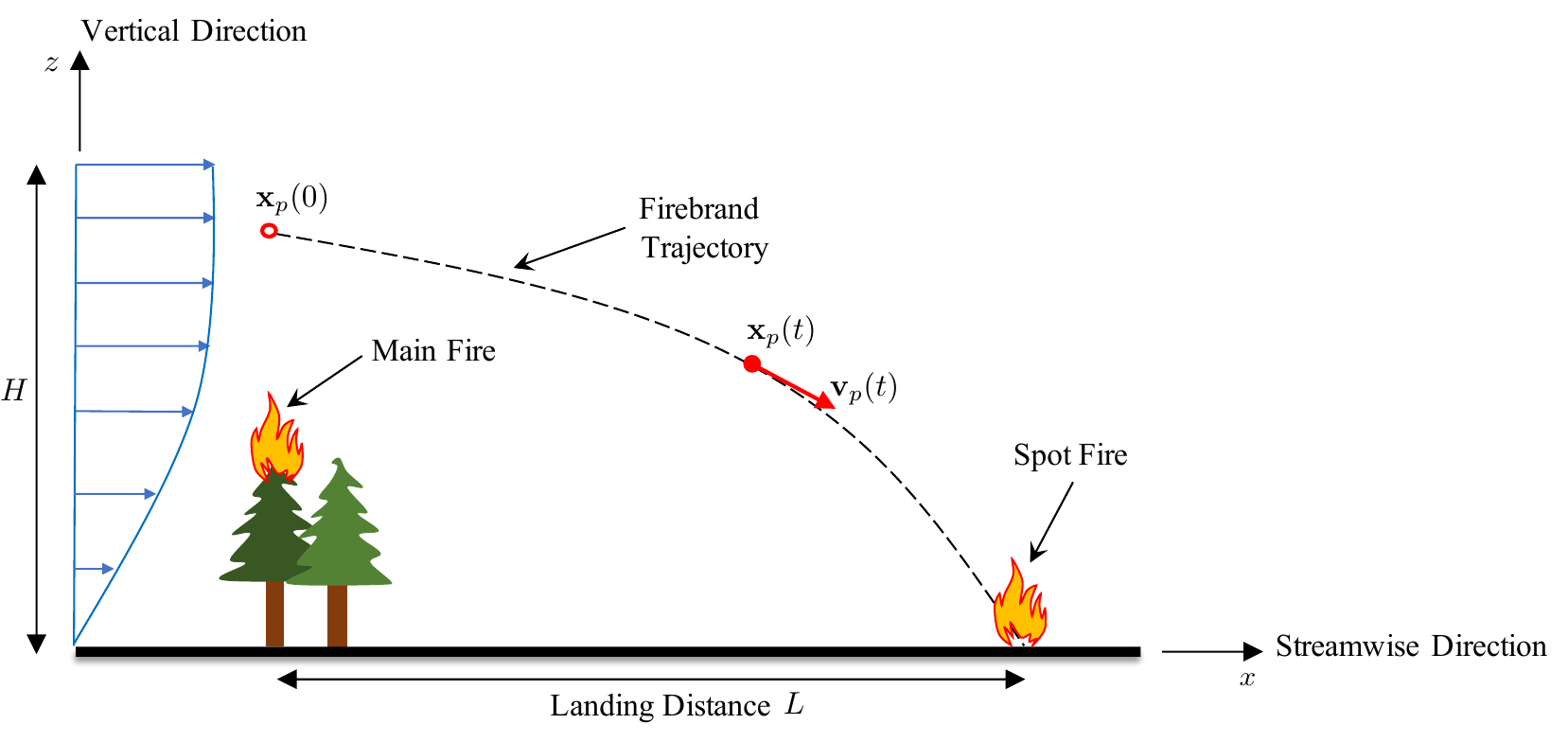}
	\caption{Schematic description of the spotting phenomenon.}
	\label{fig:schem_spotting}
\end{figure}

   Wildfires spread through two main mechanisms: (i) A diffusion-like process where fuel---such as trees, grass, and shrubs---combust in the immediate vicinity of the main fire. (ii) Spotting which refers to transport of firebrands by wind away from the main fire. At the time of landing, these firebrands may start secondary fires at a considerable distance from the main fire; see figure~\ref{fig:schem_spotting}. 
   
Forecasting spot fires remains challenging which consequently hinders our ability to effectively contain wildfires~\cite{albini1979}. For instance, New Mexico's largest wildfire started as a prescribed fire by the U.S. Forest Service in 2022 in order to safely burn the fuel on the forest floor~\cite{Nichols2024}. However, this prescribed fire quickly became out of control and evolved into a wildfire. A subsequent governmental report determined that the Forest Service underestimated the perimeter of likely spot fires and therefore did not assign personnel at a large enough distance to put them out~\cite{NM2022}.
   
Several attempts have been made to quantify the spotting distance. On the theoretical side, the seminal work of Tarifa et al.~\cite{Tarifa1963} used the governing equations of inertial particles to estimate the maximum spotting distance. A number of improvements have been made to Tarifa's estimate to account for time-dependent size of firebrands~\cite{Hellman1970}, their shape~\cite{Muraszew1976}, and the lofting process in vertical plumes~\cite{albini1979}. We refer to Section 2.6 of Ellis~\cite{Ellis2000} for a thorough review.

These theoretical estimates often underestimate the spotting distance~\cite{Muraszew1976}.
Under typical wind conditions, the theoretical estimates predict a maximum spotting distance of a few hundred meters, whereas empirical observations indicate the spotting distance can exceed one kilometer~\cite{Koo2010,Page2019}.
Moreover, the theoretical studies of maximum spotting distance often assume that the wind velocity is unidirectional, i.e., parallel to the ground. 

On the other hand, there is a wealth of computational and experimental studies that consider firebrand transport in a turbulent boundary layer~\cite{Linn2005,Kortas2009,Bhutia2010, Koo2012, Tohidi2017b,Yaghoobian2023, Yaghoobian2024}.
We refer to Wadhwani et al.~\cite{Wadhwani2022} for a comprehensive review. Although these studies are illuminating in some respects, the complexity of the wind field impedes concrete theoretical and physical insight into the spotting phenomenon.

The purpose of the present paper is to bridge the gap between the theoretical estimates of maximum spotting distance and the empirical/computational observations. To this end, we consider a class of traveling wave solutions for inviscid incompressible fluids. We first show numerically that the traveling wave can lead to spotting distances which exceed one kilometer. Then we conduct a theoretical analysis which illuminates the mechanisms responsible for such extreme spotting distances. More specifically, we first show that the firebrand motion synchronizes with the traveling wave so that the horizontal component of the firebrand velocity matches the wave speed. As a result, some firebrands---depending on their random initial condition---get trapped in a positive-lift regime of the wind. The lift force acts against gravity, leading to a surf-like motion which significantly delays the landing of the firebrand. As a result, the firebrand can travel significantly larger distances as compared to firebrand in a unidirectional wind. We derive analytical results which quantify the landing time and landing distance of firebrands in a traveling wave.

%Hellman1970: Tarifa's model but with a time-dependent radius
%
%Ellis2000 Nice review of prior work that extend Tarifa's results
%
%Koo2010 cites large spotting distance up to a few kilometers
%Page2019 observed spotting distance up to 2.7km; also indicates that Albini's max spotting distance underestimates 
%
%\citet{Kortas2009} Experimental validation of numerical models
%
%Computational studies: \citet{Bhutia2010}, Koo2012, \citet{Yaghoobian2023, Yaghoobian2024}
%
%\citet{Wadhwani2022} Review of computational results
%
%Experimental studies: \citet{Tohidi2017b}
\section{Equations of motion}
Spotting consists of three primary stages: (i) Lofting, where firebrands rise vertically in a gas plume above the main fire, (ii) Transport, where firebrands are carried downstream by the wind, and (iii) Ignition, which determines whether or not the firebrand starts a new fire after landing. This paper only focuses on the transport stage of spotting.
\subsection{Firebrand transport model}
For simplicity, we assume that the firebrands are spherical and that the wind velocity field is two-dimensional. Let $\vc x_p(t) = (x_p(t),z_p(t))$ denote the position of a firebrand at time $t$, as depicted in figure~\ref{fig:schem_spotting}. The velocity of the firebrand is given by $\vc v_p(t) = \dot{\vc x}_p(t)$, with the components $\vc v_p = (v_x,v_z)$. The trajectory of a firebrand can be obtained by solving its equations of motion~\cite{Tarifa1963,Kim1998},
	\begin{subequations}\label{eq:transp}
		\begin{equation}
		\dot{\vc x}_p = \vc v_p,
		\end{equation}
	\begin{equation}\label{eq:vdot}
	 m(t)\dot{\vc v}_p = -\frac12 \rho_fA_c C_d \|\vc v_p - \vc u(\vc x_p,t)\|\left(\vc v_p - \vc u(\vc x_p,t)\right) -m(t) \gr\vc e_z,
	\end{equation}
	\end{subequations}
	where $\|\cdot\|$ denotes  the Euclidean norm, $\rho_f$ is the fluid (air) density, $A_c=\pi r^2$ is the cross section area of the a firebrand of radius $r$, and $C_d$ is the drag coefficient. The wind velocity field $\vc u(\vc x,t)$ is two-dimensional with components $\vc u =(u_x,u_z)$. The mass of the firebrand is denoted by $m(t) = \rho_p(t) (4\pi r^3/3)$, where $\rho_p(t)$ is the mass density of the firebrand at time $t$.
	
    Equation~\eqref{eq:vdot} is Newton's second law applied to a firebrand.
    The first term on its right-hand side is the quadratic drag force and the second term is the gravitational force. With a slight abuse of terminology, we refer to the streamwise component of the aerodynamic force, $- (\rho_fA_c C_d/2) \|\vc v_p - \vc u\|(v_x-u_x)$, as drag which causes the horizontal motion of the firebrand. We refer to the vertical component, $- (\rho_fA_c C_d/2) \|\vc v_p - \vc u\|(v_z-u_z)$, as lift which counteracts the gravitational force.
    
    A few remarks about~\eqref{eq:vdot} are in order. Buoyancy, pressure gradients, and the added mass effect are orders of magnitude smaller than the aerodynamic and gravitational forces~\cite{Mendez2022}. As a result, they are safely neglected. As the firebrand burns, it loses mass which should be accounted for in the rate of change of momentum. However, it is often assumed that the mass loss takes place uniformly around the firebrand~\cite{Tarifa1963}, leading to the simplified rate of change of momentum, $m(t)\dot{\vc v}_p$.
    
    To close equation~\eqref{eq:transp}, we need to specify the time evolution of the firebrand mass $m(t)$ using a combustion model. Here, we use the empirical formula of Tarifa et al.~\cite{Tarifa1963},
    \begin{equation}
    m(t) = \frac{m_0}{1+\eta t^2},
    \end{equation}
    where $m_0=m(0)$ is the initial mass and $\eta=2.86\times 10^{-4}$ (sec$^{-2}$) is an empirical constant. We also need to specify the wind velocity field $\vc u(\vc x,t)$ which we discuss in \S\ref{sec:windModel} below.
    
    Once the appropriate initial conditions $\vc x_p(0)$ and $\vc v_p(0)$ are specified, equation~\eqref{eq:transp} can be integrated numerically to obtain the landing time $t_L$ which satisfies $z_p(t_L)=0$. The corresponding landing distance $L$ is defined as the streamwise displacement, $L = x_p(t_L)-x_p(0)$.

\subsection{Wind model}~\label{sec:windModel}
		For the wind model, we consider a family of traveling wave solutions for a two-dimensional, inviscid, and incompressible fluid. The fluid satisfies the no-flux boundary condition
		at the ground ($z=0$). Instead of the velocity-pressure formulation, it is convenient to express the governing equations in the vorticity form,
\begin{equation}\label{eq:vort}
	\pard{\zeta}{t} + \pard{\psi}{z}\pard{\zeta}{x} - \pard{\psi}{x}\pard{\zeta}{z} =0,
\end{equation}
where $\psi(x,z,t)$ is the stream function, $\zeta = -\Delta\psi$ is the vorticity field, and $\vc u = (\partial_z\psi,-\partial_x\psi)$ is the velocity field.
It is straightforward to verify that the stream function, 
\begin{equation}\label{eq:stream}
	\psi(x,z,t) = Uz+aU\sin(\ell z) \cos(kx-\omega t+\phi),
\end{equation}
is an exact solution of the vorticity equation~\eqref{eq:vort} as long as the linear dispersion relation,
\begin{equation}\label{eq:DispRel}
	\omega = kU,
\end{equation} 
is satisfied. This dispersion relation induces the wave speed $c=\omega/k=U$ which coincides with the mean wind velocity. The equality of the wave speed and the mean flow velocity will play an important role in the forthcoming analysis in \S\ref{sec:analysis}.
\begin{figure}
	\centering
	\includegraphics[width=\textwidth]{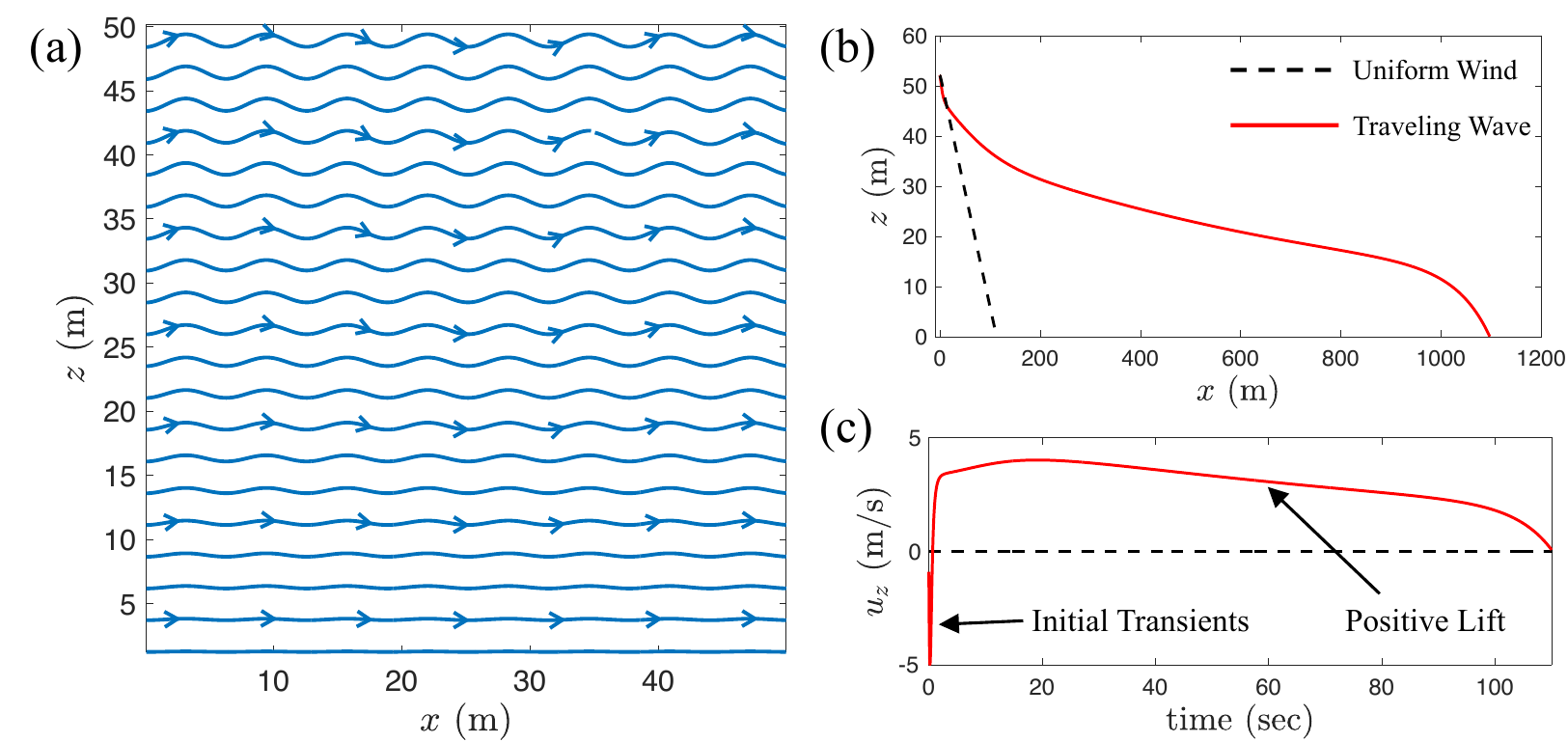}
	\caption{(a) Streamlines of the traveling wave~\eqref{eq:stream} with the mean velocity $U=10$ m/s. (b) Sample trajectories of a firebrand transported by the uniform velocity field (dashed black) and the traveling wave (solid red). In both cases, the mean velocity is $U=10$ m/s. (c) The vertical component of the wind velocity along the trajectories of the firebrands.}
	\label{fig:streamlines}
\end{figure}

We emphasize that the stream function~\eqref{eq:stream} is an exact solution of the vorticity equation for any values of the amplitude $a$, the phase $\phi$, the streamwise wavenumber $k$, the spanwise wavenumber $\ell$, and the mean velocity $U$, leading to a rich family of traveling wave solutions. The corresponding velocity field is given by 
\begin{subequations}\label{eq:tw}
	\begin{equation}\label{eq:ux}
	u_x = U + a\ell U \cos(\ell z)\cos(kx-\omega t+\phi),
	\end{equation}
	\begin{equation}\label{eq:uz}
	u_z =  a k U \sin(\ell z)\sin(kx-\omega t+\phi),
	\end{equation}
\end{subequations}
which satisfies the no-flux boundary condition at $z=0$ since $u_z(x,0,t)=-\partial_x \psi(x,0,t)=0$.
In the following, we consider two cases: (i) Uniform flow where $a=0$ and (ii) traveling waves where $a>0$.
In the case of traveling waves, we set $ak = 1/2$ so that the vertical velocity is at most one half of the mean streamwise velocity $U$. To this end, and without loss of generality, we set $a=1/2$ and $k=1$. Furthermore, we set $\ell = \pi/H$ with $H=100$ meters so that $a\ell\ll 1$ and therefore the streamwise velocity of the traveling wave is very close to that of the uniform wind. Finally, we vary $U$ between 1-10 m/s, which is the relevant range of wind velocity in the low altitudes of the atmosphere where the firebrand transport takes place~\cite{Houssami2016,Wadhwani2022}. Figure~\ref{fig:streamlines}(a) shows a snapshot of the stream function with the parameter values discussed above.

\section{Numerical results}\label{sec:numerics}
Figure~\ref{fig:streamlines}(b) shows an example of firebrand trajectories. The dashed black curve marks the trajectory of a firebrand transported by the uniform wind ($a=0$) which lands at around $x=116$ meters. The solid red curve marks the trajectory of the same firebrand transported by the traveling wave ($a=1/2$), resulting in a landing distance of approximately 1100 meters. In both cases, the mean flow velocity is $U=10$ m/s.
This figure is an example of the extreme landing distances that can be caused by traveling waves. 

Extreme landing distance under the traveling wave is in fact prevalent and not specific to this particular firebrand. In order to demonstrate this, we carry out Monte Carlo simulations and compare the landing distance distribution under uniform wind and the traveling wave~\eqref{eq:stream}. Since the initial position $(x_0,z_0)$, and the firebrand radius $r$ are uncertain, we treat them as random variables. Following the convention of denoting random variables with capital letters, these quantities are drawn from the following distributions:
\begin{equation}\label{eq:dist}
	X_0 \sim \mathcal N(0,1),\quad Z_0\sim \mathcal L\mathcal N(50,5),\quad R\sim\mathcal L\mathcal N(7.5\times 10^{-4},7.5\times 10^{-5}).
\end{equation}
Here, $\mathcal N(\mu,\sigma)$ denotes the normal distribution with mean $\mu$ and standard deviation $\sigma$. Similarly, $\mathcal L\mathcal N(\mu,\sigma)$ denotes the lognormal distribution with mean $\mu$ and standard deviation $\sigma$. Since the initial height $Z_0$ and the firebrand radius $R$ are positive, they are drawn from lognormal distributions, whereas the initial streamwise coordinate $X_0$ may assume positive or negative values and therefore is drawn from the normal distribution. All units in Eq.~\eqref{eq:dist} are in meters; for instance, the firebrand radii have a mean of 0.75\,mm and a standard deviation of 0.075\,mm meters.
This distribution is in line with the empirical results of Tohidi et al.~\cite{Tohidi2015}. The following parameters have constant values throughout this paper:
\begin{equation}
	\rho_p(0) = 513\; \mbox{kg}/\mbox{m}^{3},\quad  \rho_f=1.204\; \mbox{kg}/\mbox{m}^{3},\quad  C_d = 0.45,\quad \gr=9.8\; \mbox{m}/\mbox{s}^{2}
\end{equation}

We draw one million samples from each distribution in~\eqref{eq:dist} and integrate the transport equation~\eqref{eq:transp} for each set of random variable $(X_0,Z_0,R)$. The initial firebrand velocity $\vc v_p(0)$ is set equal to zero. As we discuss in~\S\ref{sec:analysis}, the firebrand velocity quickly converges to its asymptotic state so that the initial velocity does not have a significant impact on the results. 

\begin{figure}
	\centering
	\includegraphics[width=\textwidth]{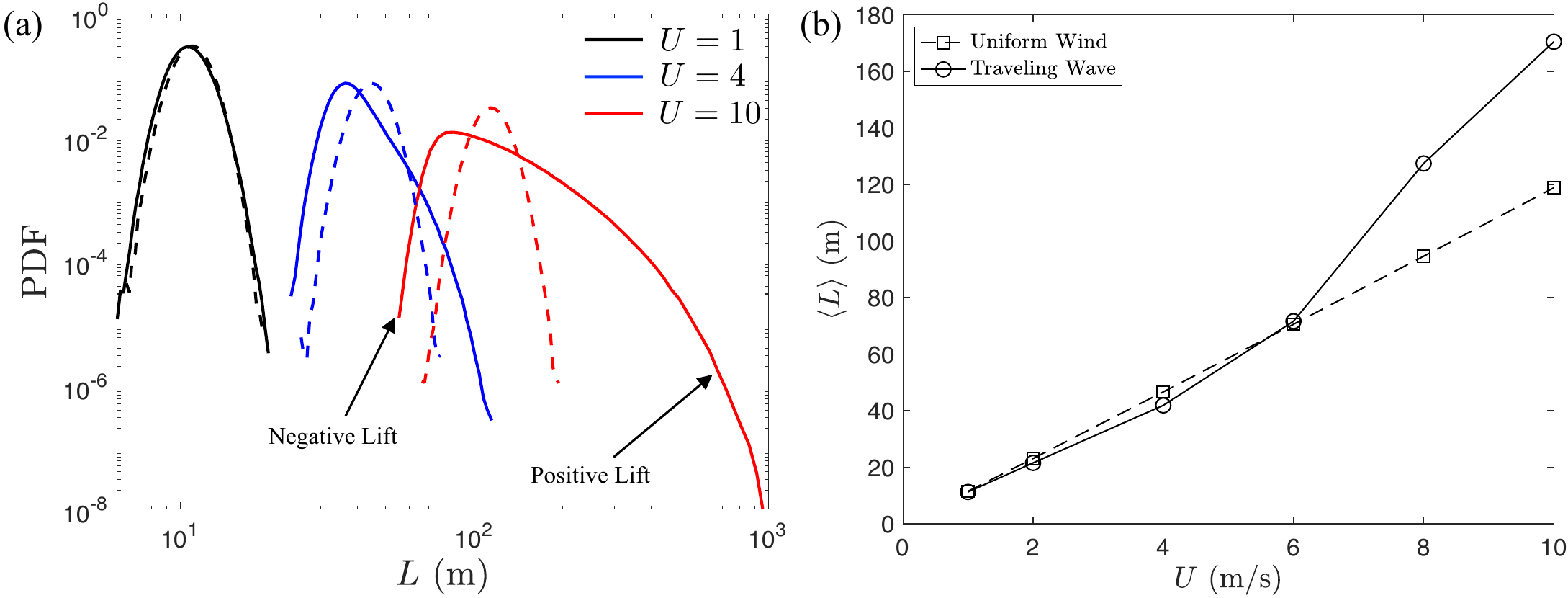}
	\caption{(a) PDF of the landing distance for three mean velocities $U=1,4,10\,$m/s. Dashed curves mark the PDF for uniform wind and solid curves correspond to traveling wave~\eqref{eq:tw}.
	(b) Mean of the landing distance $\langle L\rangle$ as a function of the mean wind speed $U$.}
	\label{fig:PDF_L}
\end{figure}

For each firebrand, we record the landing distance after numerically integrating~\eqref{eq:transp}. Note that since it depends on $(X_0,Z_0,R)$, the landing distance $L$ is itself a random variable. Figure~\ref{fig:PDF_L}(a) shows the probability density function (PDF) of the landing distance $L$ for three values of the mean velocity $U$ of the uniform wind (dashed curves) and the traveling wave (solid curves).

As expected, for both types of wind, the mean of the landing distance grows with the mean velocity; see figure~\ref{fig:PDF_L}(b) Interestingly, for the uniform wind, the distribution remains nearly Gaussian at all wind speeds. On the other hand, for the traveling wave, the landing distribution develops heavy tails as the wind speed $U$ increases. First, we note that these distributions are asymmetric around the mean. The slightly heavier tail to the left of the mean corresponds to firebrands that land more quickly compared to their counterparts which are advected by the uniform wind. The tail to the right of the mean is much heavier, indicating that the firebrands advected by the traveling wave have a significantly higher probability of landing at a larger distance.

These findings are surprising since both wind profiles---uniform and traveling wave---have the same mean streamwise velocity $U$. In \S\ref{sec:analysis}, we present a theoretical analysis which explains the heavy tails corresponding to the traveling wave as well as their asymmetric nature.

%	\subsection{Excluding other factors}
	
\section{Theory: Approximating the landing distance}\label{sec:analysis}
In this section, we show theoretically how the traveling wave can lead to much larger landing distances, in spite of the fact that its mean streamwise velocity is identical to the uniform wind field.
	
After initial transients, the firebrand quickly converges to a mechanical equilibrium where the drag force is almost zero and the lift balances gravity~\cite{Tarifa1963}. More precisely, after the initial transients, we have
\begin{align}
			\dot x_p & = v_x(t) = u_x(\vc x_p(t),t),\label{eq:xdot}\\
	        \dot z_p & = v_z(t) = u_z(\vc x_p(t), t)- \sqrt{\frac{2m(t)\gr}{\rho_fA_cC_d}},\label{eq:zdot}
\end{align}
which constitute reduced-order equations for the firebrand transport. Validity of these reduced equations have been verified against experimental and numerical results~\cite{Tarifa1963}.
Equation~\eqref{eq:xdot} implies that the horizontal displacement of the firebrand is induced by the streamwise component $u_x$ of the wind velocity. Equation~\eqref{eq:zdot}, on the other hand, implies that the vertical displacement occurs as a result of the competition between the vertical component $u_z$ of the wind and the contribution from the gravitational force. This equation is derived under the assumption that $v_z(t)\leq u_z(\vc x_p(t),t)$, which is a reasonable assumption for a typical firebrand.
	
We use reduced-order equations~\eqref{eq:xdot}-\eqref{eq:zdot} to approximate the landing time $t_L$ and the corresponding landing distance $L$. The result in~\S\ref{sec:tL_unif}, corresponding to uniform wind, were first derived by Tarifa et al.~\cite{Tarifa1963} and are presented here for completeness. Section~\S\ref{sec:tL_tw} contains our main theoretical results which quantify the landing time and distance in a traveling wave.
	
\subsection{Uniform wind}\label{sec:tL_unif}
For a uniform wind, the velocity field has no vertical component and the vertical coordinate of the firebrand can be evaluated by integrating equation~\eqref{eq:zdot} with $u_z=0$. Furthermore, it turns out that the time dependence of the firebrand mass $m(t)$ does not have a significant impact on its landing time and therefore can be set equal to the initial mass $m(t) = m_0$. With this approximation, the $z$-component of the firebrand trajectory is given by
\begin{equation}
	z(t)-z_0 = - t\sqrt{\frac{2m_0\gr}{\rho_fA_cC_d}}.
\end{equation}
At the time of landing, we have $z(t_L) =0$, and therefore
	\begin{equation}
	t_L = z_0\sqrt{\frac{\rho_fA_cC_d}{2m_0\gr}},
	\label{eq:unif_tL}
	\end{equation}
which is independent of the wind speed $U$. Since the wind velocity is uniform, $u_x = U$, the corresponding landing distance is given by
	\begin{equation}
	L = U z_0\sqrt{\frac{\rho_fA_cC_d}{2m_0\gr}} =  U z_0\sqrt{\frac{3\rho_fC_d}{8\rho_{p} r\gr}},
	\label{eq:L_unif}
	\end{equation}
which grows linearly with the wind speed $U$ and the initial height $z_0$. 
	
	Consider a typical firebrand with radius $r=0.75\,$mm released from the height $z_0=50\,$m in a uniform wind with $U=10\,$m/s. Equation~\eqref{eq:L_unif} implies a landing distance of approximately $L=116\,$m which is far smaller than kilometer-long landing distances observed in practice~\cite{Koo2010,Page2019}. For this firebrand to land at a distance of one kilometer, equation~\eqref{eq:L_unif} implies that the wind speed needs to be $U=86\,$m/s (approximately $310\,$km/h) which is unrealistic at relatively low altitudes relevant to firebrand transport.
	In \S\ref{sec:tL_tw}, we show that traveling waves, with reasonable wind speeds, can significantly increase the landing distance.
	
	\subsection{Traveling waves}\label{sec:tL_tw}
	Now we turn our attention to the firebrand transport under the traveling wave~\eqref{eq:tw}. Since $a\ell\ll 1$, the streamwise wind velocity is close to the mean flow, i.e., $u_x(\vc x,t)\simeq U$. As a result, the horizontal position of the firebrand can be approximated by $x_p(t) \simeq x_0+U t$, which is identical to that induced by a uniform wind. However, because of the non-zero vertical component of the wind, the landing time $t_L$ can differ significantly from that of a uniform wind. 
	
	The linear dispersion relation~\eqref{eq:DispRel} implies that a firebrand's streamwise velocity $U$ coincides with the wave speed $c=\omega/k$. As a result, after some initial transients, the firebrand synchronizes with the wave; i.e., it travels with the same horizontal speed. More precisely, we have $kx_p(t)-\omega t\simeq kx_0$ and therefore, the vertical component of the traveling wave velocity~\eqref{eq:uz} along the firebrand trajectory is given by $u_z(\vc x_p(t),t)\simeq A\sin(\ell z)$, where $A = akU\sin(kx_0+\phi)$ is the amplitude of the vertical component. Note that this amplitude is independent of time because the firebrand moves horizontally in sync with the wind. Therefore, depending on the random initial condition $x_0$, the firebrand will be trapped in one of the following regimes:
	\begin{enumerate}
		\item Free fall, $\sin(kx_0+\phi)\simeq 0$: In this case, the contribution of the lift force is negligible and the firebrand descends in the vertical direction only as a result of the gravitational force. This regime is similar to a firebrand moving in a unidirectional flow, such as the uniform wind.
		\item Negative lift, $\sin(kx_0+\phi)< 0$: In this case, the lift amplitude is negative, $A<0$, aligning the lift with the gravitational force. As a result, the firebrand falls more quickly compared to a unidirectional wind, leading to shorter landing distances. 
		\item Positive lift, $\sin(kx_0+\phi)>0$: This case corresponds to the a positive amplitude $A>0$ where the lift counteracts gravity, delaying the firebrand's landing. This leads to larger landing distances compared to a unidirectional wind. In the extreme case where $\sin(kx_0+\phi)\simeq 1$, the landing distance can become quite large. Figure~\ref{fig:streamlines}(c) shows the vertical component of the wind velocity along a firebrand trapped in this positive-lift regime.
	\end{enumerate}

	The remainder of this section is devoted to making the above statements quantitative. In order to approximate the landing time, we substitute the traveling wind velocity~\eqref{eq:uz} into the transport equation~\eqref{eq:zdot} to obtain,
	\begin{equation}\label{eq:zdot_tw}
		\dot z_p = A \sin(\ell z_p) - \sqrt{\frac{2m(t)\gr}{\rho_fA_cC_d}},\quad A=akU\sin(kx_0+\phi).
	\end{equation}
    If we again assume that the firebrand mass is approximately constant, $m(t) = m_0$, equation~\eqref{eq:zdot_tw} can be solved analytically. The solution has several branches depending on the initial height $z_0$. Here, we focus on the most relevant branch corresponding to $0<z_0<H$, where the exact solution is given by
    \begin{equation}
    	z_p(t) = \frac{2}{\ell}\tan^{-1} \left[\frac{A-\sqrt{G^2-A^2}\tan \left( \frac{\ell}{2}\sqrt{G^2-A^2}(t-c_0) \right)}{G}\right],
    \end{equation}
   where $\tan^{-1}$ denotes the principle value of the inverse tangent, $A = akU\sin(kx_0+\phi)$ is the amplitude of the contribution from lift, and $G = \sqrt{2m_0\gr/(\rho_fA_cC_d)}$ denotes the contribution from gravity. Solutions with $z_0>H$ can be obtained similarly from different branches of the inverse tangent function.
   The integration constant $c_0$ can be evaluated using the initial height $z_p(0) = z_0$, 
      \begin{equation}
   c_0 =\frac{2}{\ell\sqrt{G^2-A^2}}\tan^{-1}\left( \frac{G\tan(\ell z_0/2)-A}{\sqrt{G^2-A^2}}\right).
   \end{equation}
%   \begin{equation}
%   	c_0 =-\frac{2}{\ell\sqrt{G^2-A^2}}\left[ -n\pi +\tan^{-1}\left( \frac{A-G\tan(\ell z_0/2)}{\sqrt{G^2-A^2}}\right)\right],\quad \frac{n\pi}{\ell}\leq z_0<\frac{(n+1)\pi}{\ell},
%   \end{equation}
%where $n=0,1,2,\cdots$ accounts for different branches of the inverse tangent function.

\begin{figure}
	\centering
	\includegraphics[width=\textwidth]{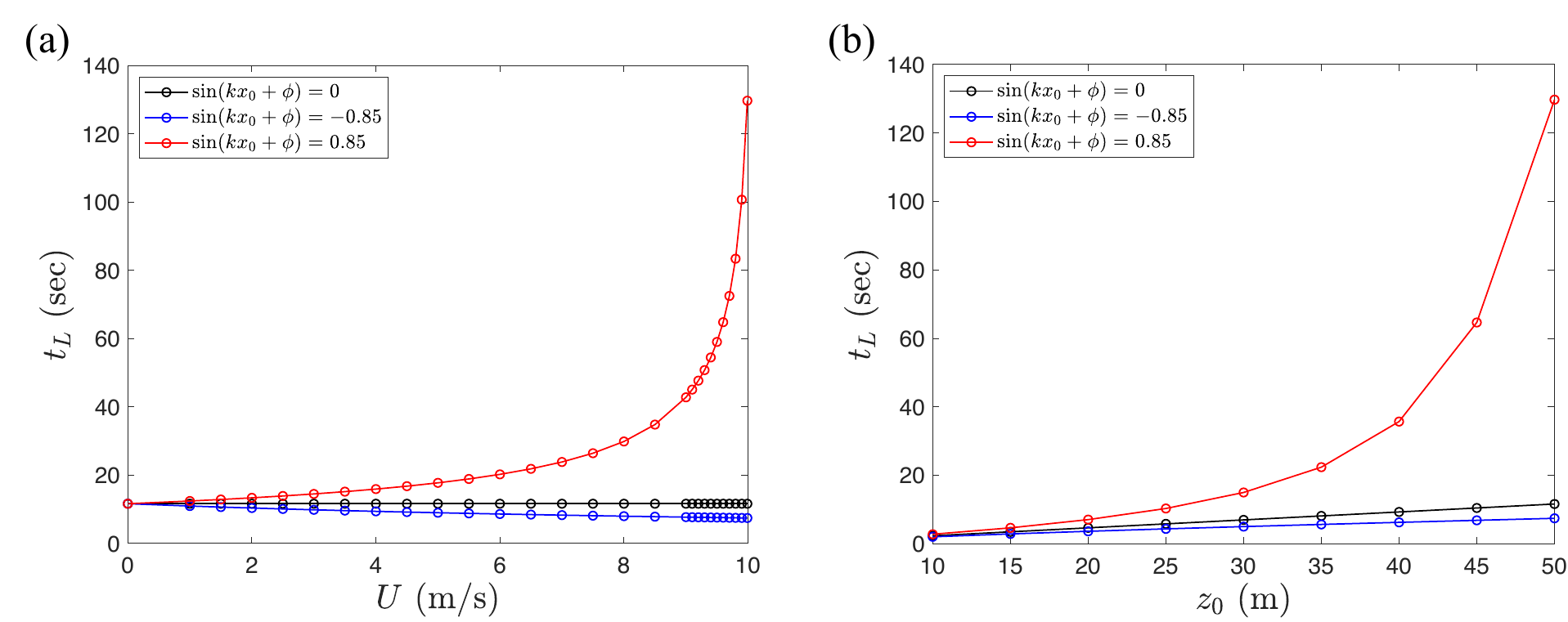}
	\caption{Theoretical landing time as a function of mean velocity and initial height. (a) As a function of the mean wind velocity $U$ with $z_0=50$ meters. (b) As a function of the initial height $z_0$ with $U=10$ m/s.}
	\label{fig:TW_tL}
\end{figure}

The corresponding landing time can be calculated using $z_p(t_L) =0$ which leads to 
\begin{equation}
	t_L = \frac{2}{\ell\sqrt{G^2-A^2}}\left[\tan^{-1}\left( \frac{A}{\sqrt{G^2-A^2}}\right)+\tan^{-1}\left( \frac{G\tan(\ell z_0/2)-A}{\sqrt{G^2-A^2}}\right)\right].
	\label{eq:TW_tL}
\end{equation}
Recall that the amplitude $A = akU\sin(kx_0+\phi)$ depends on the mean wind velocity $U$. Therefore, for a traveling wave, the landing time also depends on the mean wind velocity. This is in contrast to the landing time~\eqref{eq:unif_tL} for the uniform wind which is independent of the mean velocity. 

Figure~\ref{fig:TW_tL}(a) shows the landing time~\eqref{eq:TW_tL} as a function of the mean velocity $U$ for a firebrand of radius $0.75\,$mm released from the initial height $z_0=50$ meters.
In the free-fall regime where $\sin(kx_0+\phi) =0$, the landing time is independent of the mean velocity, retrieving the uniform wind results. In the negative-lift regime where $\sin(kx_0+\phi) =-0.85$, the landing time decreases slightly compared to the free-fall regime. Most interestingly, in the positive-lift regime where $\sin(kx_0+\phi) =0.85$, the landing time increases significantly with the mean velocity. In particular, as $U$ approaches 10 m/s, the landing time reaches 130 seconds. This implies a landing distance of $L \simeq  Ut_L=1300$ meters. Contrast this with the same firebrand in the uniform wind which travels a distance of $L\simeq 116$ meters before landing.

A similar behavior can be observed for the landing time as a function of the initial height $z_0$. Figure~\ref{fig:TW_tL}(b) shows this relationship for $U=10$ m/s.
In the free-fall regime, the landing time increases linearly with the initial height, similar to the uniform wind. In the negative-lift regime, the landing time also increases almost linearly with the initial height but at a slightly lower rate. On the other hand, in the positive-lift regime, the landing time increases significantly with the initial height.

These results explain the asymmetric heavy tails observed in figure~\ref{fig:PDF_L}(a). Namely, firebrands trapped in a negative-lift regime ($A<0$) land faster than a firebrand in a uniform wind. However, their landing time is only slightly smaller than a firebrand in uniform wind; see figure~\ref{fig:TW_tL}. This leads to slightly heavier tails to the left of the mean in figure~\ref{fig:PDF_L}(a).
On the other hand, firebrands trapped in the positive-lift regime ($A>0$) have a landing time significantly larger than those in the uniform wind. This leads to the significantly heavier tails to the right of the mean.

In closing, we emphasize that our numerical results in \S\ref{sec:numerics} where obtained from the full-order model~\eqref{eq:transp}, whereas our theoretical results in this section were derived from the reduced-order model~\eqref{eq:xdot}-\eqref{eq:zdot}. Nonetheless, our theoretical findings yielded significant insight into the behavior of firebrands in a traveling wave.
	
\section{Conclusions}
We showed that atmospheric traveling waves can significantly increase the spotting distance in wildfires. Our theoretical analysis revealed the physical mechanism underlying this phenomenon: Firebrands synchronize with the traveling wave such that their horizontal velocity matches the wave speed. As a result, some firebrands---based on their random initial conditions---get trapped in a positive-lift region of the wind and experience a surf-like motion. This delays the firebrand's landing and leads to significantly larger spotting distances compared to firebrands transported by unidirectional winds.

The relatively simple form of the monochromatic traveling wave~\eqref{eq:stream} allowed for our detailed theoretical analysis.
It remains to be seen if similar synchronization and trapping mechanisms hold for more complex wind conditions, such as lee waves and turbulent boundary layers.
The main impediment to generalizing our theory to turbulent boundary layers is the validity of the reduced-order equations~\eqref{eq:xdot}-\eqref{eq:zdot} which are leading-order approximations. Higher-order approximations have been derived for low Reynolds number flows where the drag force is linear~\cite{MR,rubin1995_IP,IP_haller08}. At the moment, such more accurate reduced-order equations are missing for the quadratic drag~\eqref{eq:vdot} which is the correct model at high Reynolds numbers relevant to firebrand transport.

\paragraph{Acknowledgments.}This work was supported by the National Science Foundation, the Algorithms for Threat De- tection (ATD) program, through the award DMS-2220548.
	
%\bibliographystyle{plain}
%\bibliography{../../../bibliog.bib}

\end{document}